\documentclass[%
twocolumn,
showpacs,preprintnumbers,
nofootinbib,
nobibnotes,
bibnotes,
amsmath,amssymb,
aps,
prd,
]{revtex4}

\usepackage{graphicx}
\usepackage{dcolumn}
\usepackage{bm}

\begin{document}

\title{Dynamic analysis of noncanonical warm inflation}

\author{Xi-Bin Li}
\email{lxbbnu@mail.bnu.edu.cn}
\affiliation{Department of Physics, Beijing Normal University, Beijing 100875, China}

\author{Yang-Yang Wang}
\email{hewang@mail.bnu.edu.cn}
\affiliation{Department of Physics, Beijing Normal University, Beijing 100875, China}

\author{He Wang}
\email{wangyy@mail.bnu.edu.cn}
\affiliation{Department of Physics, Beijing Normal University, Beijing 100875, China}

\author{Jian-Yang Zhu}
\thanks{Corresponding author}
\email{zhujy@bnu.edu.cn}
\affiliation{Department of Physics, Beijing Normal University, Beijing 100875, China}

\date{\today}

\begin{abstract}
We study and analyze the dynamic properties of both canonical and noncanonical warm inflationary models with dissipative effects. We consider different models of canonical warm inflation with different dissipative coefficients and prove that the behavior at infinity of quadratic dissipative model distinctly differs from that of the constant dissipative model, which means that quadratic dissipative coefficient increases the possibility of the occurrence of inflation. We also show that the different choice of combination of the parameters in noncanonical warm inflation exhibits dramatically different global phase portraits on the Poincar\'{e} disk. We try to illustrate that the noncanonical field will not expand the regime of inflation, but it will increase the possibility of the occurrence of inflation as well and the duration of inflation. Then, by dynamic analysis, we can exclude several inflationary models, like the warm inflation model, with negative dissipative coefficients, and explain that the model without potential is almost impossible. With relevant results, we give the condition when reheating occurs.

\end{abstract}

\maketitle


\section{\label{introduction}Introduction}

Inflation is an extremely successful model that provides a graceful method to overcome the shortcomings of the standard cosmological model, like the horizon problem and the flatness problem \cite{RevModPhys.78.537}, which is consistent with the cosmological observations of the large-scale CMB \cite{2016A&A...594A..13P, 2014A&A...571A..16P} and large-scale structure \cite{PhysRevD.74.123507,doi:10.1111/j.1365-2966.2005.09318.x}. In this simplest and elegant model, the inflation can be described by a special period when our Universe expands rapidly driven by a nearly constant energy density arising from the potential of a scalar field \cite{2003moco.book.....D,RevModPhys.57.1}.

With the success of inflationary scenario, series of candidate scenarios have been established, among which warm inflation is a model that reckons the early Universe has a moderate temperature instead of being cold \cite{RevModPhys.78.537,PhysRevD.62.083517}. From the point of view of this scenario, particles interact with other fields and decay to other particles, which leads to an effect of a friction term to describe this decay phenomenon during inflation \cite{PhysRevD.91.083540,PhysRevD.71.023513}. As a result, the primary source of density fluctuations comes from thermal fluctuations \cite{PhysRevD.64.063513,PhysRevD.69.083525} rather than from quantum fluctuations. However, it was realized a few years later that the idea of warm inflation was not easy to realize in concrete models and, even simply, not possible \cite{PhysRevD.60.083509,PhysRevLett.83.264}. Soon after, successful models of warm inflation have been established, in which the inflaton indirectly interacts with the light degrees of freedom through a heavy mediator field instead of being coupled with a light field directly \cite{PhysRevD.84.103503,PhysRevLett.117.151301,1475-7516-2011-09-033}. The warm inflation model has been widely studied by a series of methods, like field theory method \cite{PhysRevD.50.2441} and stability analysis \cite{PhysRevD.90.123519,1475-7516-2012-11-042}. Dynamic analysis is also an effective method to analyze the dynamical properties of warm inflationary system \cite{PhysRevD.57.741,PhysRevD.73.023502,1475-7516-2010-08-002}.

Another simple way to establish the inflationary scenario involves extending the Lagrangian density from a canonical kinetic term to a noncanonical one \cite{1475-7516-2012-08-018,PhysRevD.78.023517}. The noncanonical inflationary scenario has some interesting features, such as that the equations of motion remain second order and that the slow-roll conditions become easier to realize compared to canonical inflationary theory \cite{1475-7516-2012-08-018}. Most noncanonical models can drop the tensor-to-scalar ratio considerably \cite{PhysRevD.83.101301}, and stability analysis shows that many such models have stable attractors \cite{PhysRevD.81.123526}. The work under the frame of noncanonical inflation has been done numerous times \cite{1475-7516-2013-03-028,1475-7516-2013-12-010,ARMENDARIZPICON1999209,1475-7516-2006-02-004}. Recently, relevant researches have shown that noncanonical warm inflationary models still satisfy the stability condition as long as each model controls parameters at a moderate stage \cite{PhysRevD.90.123519}, based on which noncanonical inflationary models are being extended to warm scenario, such as the warm-DBI model \cite{PhysRevD.83.101301}, warm $k$-inflation \cite{PhysRevD.94.103531,PhysRevD.97.063523}, and so on \cite{PhysRevD.87.043522,1475-7516-2014-02-005}.
l
In this paper, we attempt to illustrate the global dynamic behaviors of both canonical and noncanonical warm inflation on planar phase space. We start from the condition of canonical inflation and obtain its singularities at the original point and infinite region in the presence of different dissipative coefficients. Based on the result from the canonical one, we further study the global dynamical behaviors of warm inflationary models with different Lagrangian density of noncanonical field. Then we extend our work to some models with more complex topological structures in phase space.

The paper is organized as follows. In Sec. \ref{system}, we derive the basic dynamic equations from relevant physical equations and assumptions and define the inflationary region in phase space that applies to our study. In Sec. \ref{canonical_inflation}, we study the canonical warm inflationary models with a constant dissipative coefficient and a quadratic field dependent dissipative coefficient by a mathematical method. In Sec. \ref{Noncanonical_Warm_inflation},  we focus on the global dynamic phase portraits for different noncanonical warm inflation and get series of interesting and inspired results. In Sec. \ref{other}, we study some models with some strange topological structures like limit cycle and Hopf bifurcation, and use them to exclude some inflationary scenarios and discuss when reheating could be realized. In Sec. \ref{conclusion}, conclusions and relevant further discussions are given.

\section{\label{system}The Differential System}

The action of noncanonical warm inflation writes
\begin{eqnarray}
     S = \int{d^4x \sqrt{-g}[\mathcal{L}_{\textrm{non-con}}(X,\phi)+\mathcal{L}_{\textrm{R}}+\mathcal{L}_{\textrm{int}}]}, \label{total_action}
\end{eqnarray}
where $X=\frac{1}{2}g^{\mu\nu}\partial_\mu\phi\partial_\nu\phi$, $\mathcal{L}_{\textrm{non-con}}(X,\phi)$ is the noncanonical Lagrangian density of field, $\mathcal{L}_{\textrm{R}}$ is the Lagrangian density of radiation field, $\mathcal{L}_{\textrm{int}}$ is the Lagrangian density of interaction between inflaton and other fields, and $g$ is the determinant of metric $g_{\mu\nu}=\textrm{diag}(-1,a^2(t),a^2(t),a^2(t))$. The null energy condition and the physical propagation of perturbations require that $\mathcal{L}_X\geqslant0$ and $\mathcal{L}_{XX}\geqslant0$ \cite{PhysRevD.81.123526,PhysRevD.78.023517}, and the subscript $X$ denotes a derivative with respect to $X$. The equation of motion can be obtained by taking the variation of the action:
\begin{eqnarray}
     \bigg[\frac{\partial\mathcal{L}(X,\phi)}{\partial X}+2X\frac{\partial^2\mathcal{L}(X,\phi)}{\partial X^2}\bigg]\ddot{\phi}\nonumber \\
     +\bigg[3H\frac{\partial\mathcal{L}(X,\phi)}{\partial X}+\dot{\phi}\frac{\partial^2\mathcal{L}(X,\phi)}{\partial X \partial \phi}\bigg]\dot{\phi}\nonumber \\
     -\frac{\partial\mathcal{L}(X,\phi)}{\partial \phi}=0, \label{EOMnc}
\end{eqnarray}
where $H\equiv \dot{a}/a$ denotes the Hubble parameter, and a dot means a derivative with respect to the cosmic time $t$. The Lagrangian density of the noncanonical field can be writen in a simple form as \cite{ARMENDARIZPICON1999209}
\begin{eqnarray}
     \mathcal{L}_{\textrm{non-con}}(X,\phi)=K(\phi)X+\alpha X^2-V(\phi), \label{noncanonical_action}
\end{eqnarray}
where $K(\phi)$ is called "kinetic function" and $V(\phi)$ is the potential function of $\phi$. The energy-momentum tensor is written as $T_{\mu\nu}=(\partial\mathcal{L}/\partial X)\partial_\mu\phi\partial_\nu\phi-g_{\mu\nu}\mathcal{L}$. Thus the energy density and pressure are, respectively
\begin{eqnarray}
     & &\rho_\phi = K(\phi)X+3\alpha X^2+V(\phi), \label{energy_density}\\
     & &p_\phi = K(\phi)X+\alpha X^2-V(\phi). \label{pressure}
\end{eqnarray}
In the warm inflation model, there is a dissipation term to describe the inflaton fields coupling with the thermal bath. With this assumption and Eq. (\ref{noncanonical_action}), we obtain \cite{PhysRevD.94.103531}
\begin{eqnarray}
     (3\alpha \dot{\phi}^2+K)\ddot{\phi}+3H(\alpha\dot{\phi}^2+K)\dot{\phi}+\Gamma\dot{\phi}+\frac{1}{2}K_\phi\dot{\phi}^2+V_\phi=0, \nonumber\\ \label{equation}
\end{eqnarray}
where $\Gamma(\phi)$ is the dissipative term.

In order to get the complete differential dynamic equation, we also need two Einstein equations,
\begin{eqnarray}
     H^2=\frac{8\pi G}{3}(\rho_\phi+\rho_\textrm{R})-\frac{k}{a^2}, \label{Einstein_equation1}\\
     2\dot{H}+3H^2+\frac{k}{a^2}=-8\pi G(p_\phi+p_\textrm{R}), \label{Einstein_equation2}
\end{eqnarray}
where $p_\textrm{r}=\frac{1}{3}\rho_\textrm{R}$ is the pressure of radiation field. In this paper, we consider only the condition of  homogeneous and flat spacetime, i.e., $k=0$ and $X=\frac{1}{2}\dot{\phi}^2$. We have that $\rho_\phi$ and $\rho_\textrm{R}$ evolve in time as \cite{0034-4885-72-2-026901}
\begin{eqnarray}
     \dot{\rho}_\phi+3H(\rho_\phi+p_\phi)+\Gamma(\phi)\dot{\phi}^2=0, \label{rho_phi}\\
     \dot{\rho}_\textrm{R}+4H\rho_\textrm{R}-\Gamma(\phi)\dot{\phi}^2=0. \label{rho_R}
\end{eqnarray}
From Eq. (\ref{Einstein_equation1}), We can consider
\begin{eqnarray}
      \rho_\textrm{R}& &=\frac{3}{8\pi G}H^2-\rho_\phi \nonumber \\ & &=\frac{3}{8\pi G}H^2-K(\phi)X-3\alpha X^2-V(\phi) \label{E_rho_R}
\end{eqnarray}
as the expression of $\rho_\phi$ in Eq. (\ref{rho_R}). Thus, we can write $\dot{H}$ as
\begin{eqnarray}
      \dot{H}=-2H^2-\frac{8\pi G}{3}(K(\phi)X-2V(\phi)), \label{E_H}
\end{eqnarray}
where we have used $p_\phi$ and $\rho_\phi$ in Eqs. (\ref{pressure}) and (\ref{E_rho_R}).

Now, from Eqs. (\ref{equation}) and (\ref{E_H}), together with $\dot{\phi}=d\phi/dt$, we get a three-dimensional dynamic system in phase space of $(\phi,\dot{\phi},H)$. After dimensionless treatment, the differential system becomes
\begin{widetext}
\begin{gather}
        \dot{x}=y, \nonumber \\
        \big(K(x)+\alpha y^2\big)\dot{y}=-3\big(K(x)+\alpha y^2\big)yz-\Gamma(x)-\frac{1}{2}K_xy^2-V_x,\label{Dyn_sys_nd}\\
        \dot{z}=-2z^2-\frac{8\pi G}{3}\big(\frac{1}{2}K(x)y^2-2V(x)\big). \nonumber
\end{gather}
\end{widetext}
It is noteworthy that any variable or parameter in the equations above is dimensionless and its physical meaning will be introduced in next sections.

The physical region in three-dimensional phase space is defined by the condition $\rho_\textrm{R}\geqslant 0$. In general, the dissipative coefficient is not arbitrary but with the form \cite{PhysRevD.57.741}
\begin{eqnarray}
      \Gamma(\phi)=\Gamma_n\phi^n, \label{dissipative_coefficient}
\end{eqnarray}
where $n$ is an even number. In this condition, those trajectories that initially lie inside the region $\rho_\textrm{R}\geqslant 0$ remain holding in this region, in which the trajectories will neither cross the region $\rho_\textrm{R}=0$ nor enter the region $\rho_\textrm{R}\leqslant 0$. We will also show in this paper that any parameter in $\mathcal{L}_{\textrm{non-con}}(X,\phi)$ or $\Gamma(\phi)$ must be a positive number or an odd exponential functional form. These forms of relevant functions provide the singularities at infinity is symmetry about the original point, when all trajectories lie in phase planar will not cross the infinite boundary but keep inside the Poincar\'{e} disk. The inflationary region is defined by
\begin{eqnarray}
      \frac{\ddot{a}}{a}=\dot{H}+H^2>0 \label{inflationary_region}
\end{eqnarray}
labeled as $\mathcal{J}$, which must be located in the region with positive curvature,
\begin{eqnarray}
      R=6(\dot{H}+2H^2)>0 \label{positive_curvature_region}
\end{eqnarray}
labeled as $\mathcal{R}$. Generally the analytical form of $\mathcal{J}$ is quite complex, but sometimes there is no need to know the exact formula. On the contrary, we can plot it on an approximate region located in $\mathcal{R}$ which is tangential with $\mathcal{J}$ at infinity. This approximate method is widely used in Sec. \ref{Noncanonical_Warm_inflation}.

We attempt, however, to discuss the dynamic system in two-dimensional phase space $(x,y)$ (consider $z$ as a constant) instead of three-dimensional. The reasons are as follows:
\begin{itemize}
  \item The trajectory of $z$ is quite simple in which it is just a monotone decreasing curve, so the trajectory in three-dimensional phase space $(x,y,z)$ is the topological equivalent to the trajectory in two-dimensional phase space $(x,y)$.
  \item Numerical analysis shows that the behavior of $z$ trajectories evolves very slowly, so the topological structures of them are almost the same.
  \item What we are interested in most is the slow-roll condition; i.e., $\varepsilon\equiv-\dot{H}/H^2\ll 1$, which is consistent with the reason just above.
  \item The dynamic system in Eq. (\ref{Dyn_sys_nd}) domains by variables $x$ and $y$, and the presence of $z$ will not generate any complicated structures, like chaos or singular closed trajectory.
\end{itemize}

However, there are two points that we need to point out. First, the approximation above is based on the mathematical aspect. Physically, the invariance of the Hubble parameter means the conservation of entropy, including that of matter and fields on the cosmic horizon together with those inside the horizon\cite{arxiv1801.06155,Herrera2018}. But in cosmic settings (with the absence of a black hole), it has been formulated that said entropy would not diminish. An ordinary way to solve this problem is by experiencing the Hubble parameter as a slight decrease function, i.e., $H(t)=\bar{H}(1-\epsilon(\phi))$, where $\epsilon(\phi)$ is a slight increase function with value much smaller than the unit during $[0,\infty)$. Now, the existence of $\epsilon$ will make the differential dynamic system (\ref{Dyn_sys_nd}) a little more complicated, but, as illustrated above, the existence of such a slight and small valued function will not change the topological structure on the Poincar\'{e} disk compared with the original dynamic system. So it is enough for the case of the constant Hubble parameter to explain the global dynamical properties on phase space. Secondly, the dissipative coefficient $\Gamma$ in Eq. (\ref{dissipative_coefficient}) is a function only dependent on inflaton field $\phi$. But, generally, it depends on both inflaton field $\phi$ and temperature $T$, as an example \cite{1475-7516-2009-03-023} but not established,
\begin{eqnarray}
      \Gamma(\phi)=\Gamma_n\frac{\phi^{2n}}{T^{2n-1}}. \label{dissipative_coefficient'}
\end{eqnarray}
If we consider $T\propto a^{-1}=e^{-Ht}$, the differential dynamic system of Eq.(\ref{Dyn_sys_nd}) is a nonautonomous system in which the vector field $Q(x,y,t)$ (see the Appendix \ref{definitions_and_properties})consists of times explicitly. Also, the topological structure of the nonautonomous system is the same with the situation of autonomous one, since the time dependent function $e^{Ht}$ has no singularity on the duration $[0,\infty)$ \cite{2619394820151001,kloeden2011nonautonomous}. So, to be convenient, we only concentrate on the condition that the dissipative coefficient $\Gamma$ is independent on $T$ as shown in Eq. (\ref{dissipative_coefficient}). This causes us, however, to study the dynamical system with the temperature dependent dissipative coefficient $\Gamma=\Gamma(\phi,T)$ together with the temperature dependent effective potential $V(\phi,T)$. In a word, in this paper, we focus on the possibility and extent that the trajectories cross through the inflationary region $\mathcal{J}$, and it is enough to illustrate these two crucial problems with the conditions of the temperature independent dissipative coefficient and the constant Hubble parameter.

Next we will show several examples in two-dimensional phase space in different models.

\section{\label{canonical_inflation}Dynamic analysis of canonical inflation}            

\begin{figure}
  \centering
  \includegraphics[width=3.0in,height=3in]{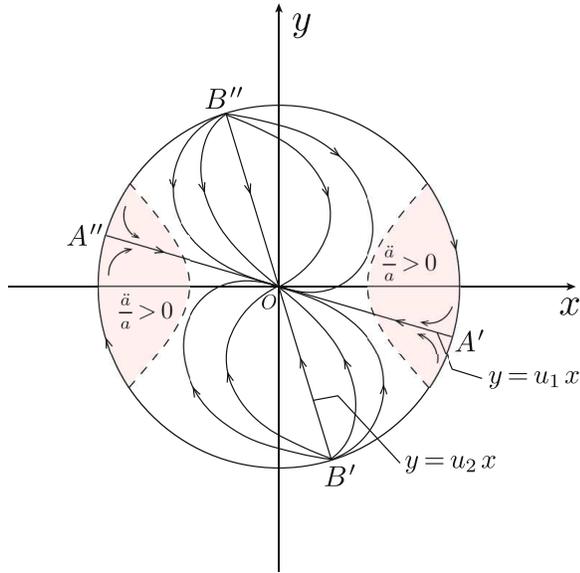}
  \caption{Global phase portrait of the dynamic system in Eq. (\ref{Dyn_sys_gamma0}).}\label{canonical_gamma0}
\end{figure}

Before analyzing the dynamic properties of noncanonical warm inflation, we first consider the canonical one, which will help us better understand the properties of both.

\subsection{\label{Gamma_constant}$\Gamma(\phi)=\textrm{constant}$}                     
\begin{figure}
  \centering
  \includegraphics[width=3.0in,height=3in]{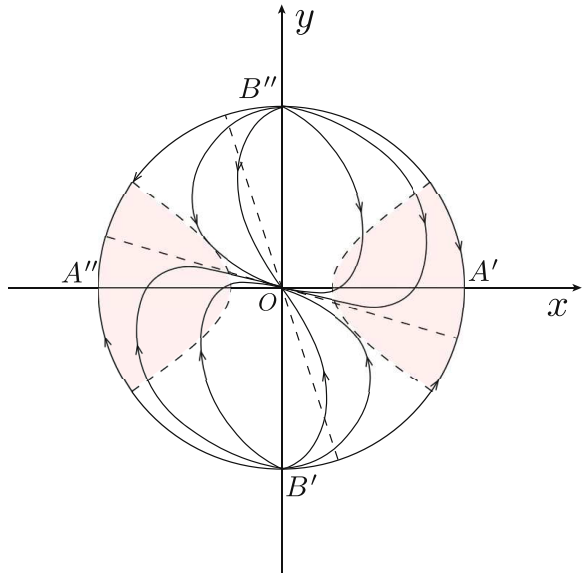}
  \caption{Global phase portrait of the dynamic system in Eq. (\ref{Dyn_sys_gamma2}).}\label{canonical_gamma2}
\end{figure}
\begin{figure*}
  \centering
  \includegraphics[width=6.0in,height=2.7in]{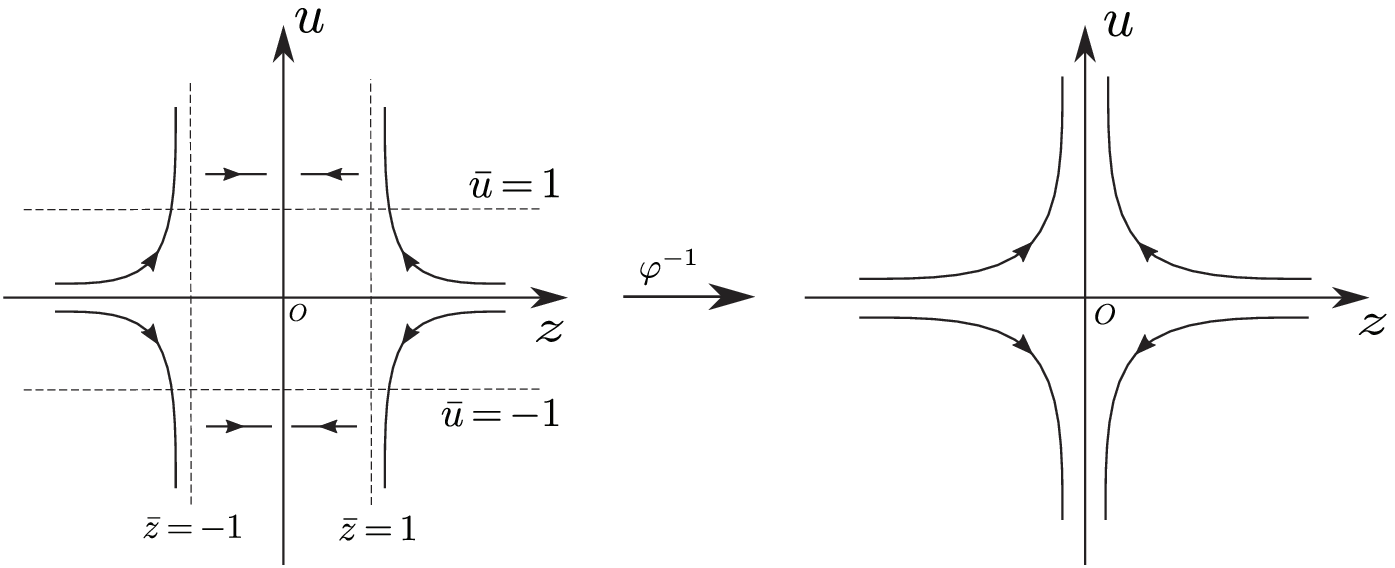}
  \caption{Blowup of the singularity and local phase portrait of Eq. (\ref{Dyn_sys_gamma2_infinity}).}\label{blowup_canonical}
\end{figure*}
Let us start from a simple condition. As an easy example, we first consider the canonical warm inflation with a constant dissipation coefficient. Set the following: $K(\phi)=1$, $\alpha=0$, $\Gamma(\phi)=\Gamma_0$, and $V(\phi)=\frac{1}{2}m^2\phi^2$. Redefine the variables $t\rightarrow t/m$, $\phi\rightarrow M_px$, $\dot{\phi}\rightarrow mM_py$, $H\rightarrow m\bar{H}$, and $\Gamma_0\rightarrow m\bar{\Gamma}_0$. The expression of such a dynamic dissipative system is quite simple:
\begin{eqnarray} \label{Dyn_sys_gamma0}
      \begin{cases}
        \dot{x}=y,  \\
        \dot{y}=-x-(3\bar{H}+\bar{\Gamma}_0)y.
      \end{cases}
\end{eqnarray}
To be convenient, we set $r\equiv 3\bar{H}+\bar{\Gamma}_0\gg 1$.

First, we need to research the topological structure at a singular point $(0,0)$. The stability topological structure at $(0,0)$ is determined by the matrix
\begin{eqnarray}
      \textbf{A}=\begin{pmatrix}
        0 & 1\\
        -1 & -r \end{pmatrix}, \label{A1}
\end{eqnarray}
whose eigenvalues are $\lambda_1=k_1\ll -1$ and $\lambda_2=k_2\lesssim 0$, with $k_1k_2=1$.
\footnote{
The system in Eq. (\ref{Dyn_sys_gamma0}) satisfies the differential equation
\begin{eqnarray}
      \frac{dy}{dx}=\frac{-x-ry}{y}=\frac{-1-r(y/x)}{y/x}. \nonumber
\end{eqnarray}
Set $k=y/x$ at $(0,0)$; we have
\begin{eqnarray}
      k^2+rk+1=0, \nonumber
\end{eqnarray}
with roots
\begin{eqnarray}
      \lambda_1=k_1=\frac{-r-\sqrt{r^2-4}}{2}\ \textrm{and}\ \lambda_2=k_2=\frac{-r+\sqrt{r^2-4}}{2}.  \nonumber
\end{eqnarray}}
According to the definition \ref{definnition_of_elementary_singularity} in Appendix\ref{definitions_and_properties}, the singular point $(0,0)$ is a stable node with a topological structure like the first pattern in Fig. \ref{tological_singularity}(see Appendix).

Next we analyze the singularity at infinity. To study the orbits which tend to or come from infinity, we can apply Poincar\'{e} compactification \cite{Dynamic1}, which will tell us the topology in the infinite region. Do the coordinate transformation
\begin{eqnarray}
     \phi:(x,y)\mapsto(u,z)=\left(\frac{y}{x},\frac{1}{x}\right). \label{Poincare}
\end{eqnarray}
This transformation maps the point on infinity to $\mathbb{S}^2$. Then the system in Eq. (\ref{Dyn_sys_gamma0}) is given by
\footnote{
Use the relation
\begin{eqnarray}
      \begin{cases}
        \dot{u}=\frac{1}{x}\dot{y}-\frac{x}{y^2}\dot{x}=zQ\Big(\frac{1}{z},\frac{u}{z}\Big)-uzP\Big(\frac{1}{z},\frac{u}{z}\Big), \nonumber \\
        \dot{z}=-\frac{1}{x^2}\dot{x}=-z^2P\Big(\frac{1}{z},\frac{u}{z}\Big). \nonumber
      \end{cases}
\end{eqnarray}} 
\begin{eqnarray} \label{Dyn_sys_gamma0_infinity}
      \begin{cases}
        \dot{u}=-(u^2+ru+1), \\
        \dot{z}=-zu.
      \end{cases}
\end{eqnarray}
Set $z=0$ (that means the singular point at infinity), then we get two singularities $(u_1,0)$ and $(u_2,0)$ with $u_1=k_1$ and $u_2=k_2$. This result tells us that there are four singularities, which are distributed along the directions $y=u_1x$ and $y=u_2x$. By studying the stability, it is easy to find that singular points $A'$ and $A''$ on $y=u_2x$ are saddles which are symmetric with the original point of the planner $(x,y)$, while singular points $B'$ and $B''$ on $y=u_1x$ are unstable nodes which are also symmetric with the original point (see Fig. \ref{canonical_gamma0}).

The dynamic system has no more singularities on $\mathbb{R}^2$. If we plot the singularities above on one finite plane, which is also called the Poincar\'{e} disk, we obtain the global phase portrait (also plotted on the panel in Fig. \ref{canonical_gamma0}). The portrait shows that the directions $y=u_1x$ and $y=u_2x$ are not equivalent, i.e., that $y=u_2x$ is more stable than $y=u_1x$. So the direction along $y=u_2x$ is a strong direction while $y=u_1x$ is called a weak direction. Detailed calculations show $\{(x,y)|y=u_2x\}\bigcap\mathcal{J}\neq \emptyset$, which means that most trajectories will cross the inflationary region.

\subsection{\label{Gamma_quadric}$\Gamma(\phi)=\Gamma_2 \phi^2$}                    
\begin{figure*}
  \centering
  \includegraphics[width=6.0in,height=2.7in]{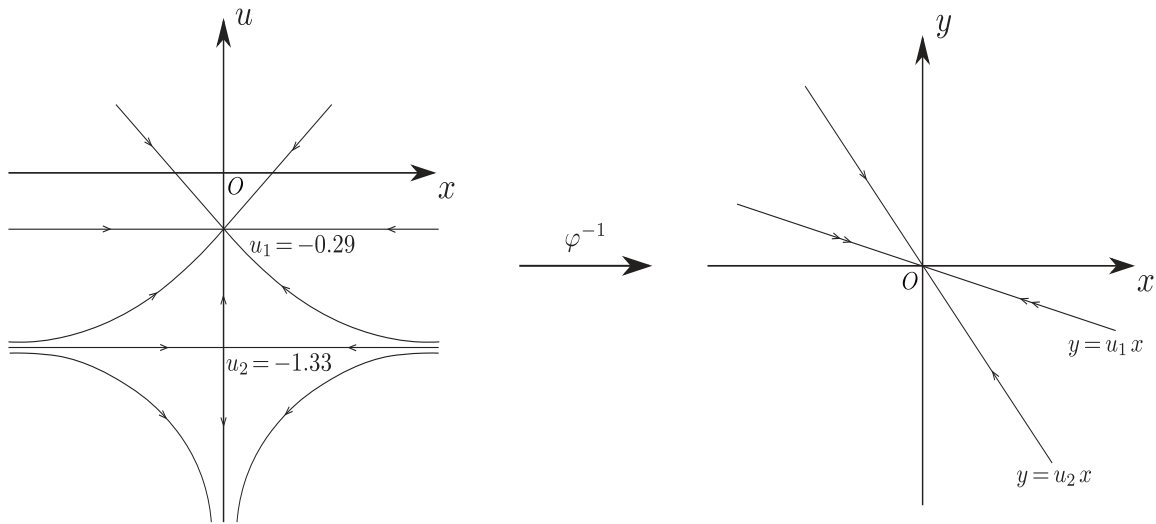}
  \caption{Blowup of the singularity and local phase portrait of Eq. (\ref{blowup_case1}).}\label{blowup_noncanonical_1}
\end{figure*}

Let us start from a simple condition. As an easy example, we first consider the canonical inflation. Now, set the following: $K(\phi)=1$, $\alpha=0$, $\Gamma(\phi)=\Gamma_2\phi^2$ and $V(\phi)=\frac{1}{2}m^2\phi^2$. Redefine the variables $t\rightarrow t/m$, $\phi\rightarrow M_px$, $\dot{\phi}\rightarrow mM_py$, $H\rightarrow m\bar{H}$, and $\Gamma_2\rightarrow (m/M_p^2)\bar{\Gamma}_2$; we obtain the dynamic dissipative system:
\begin{gather}
      \begin{cases}
        \dot{x}=y,  \\
        \dot{y}=-x-3\bar{H}y-\Gamma_2x^2y.
      \end{cases} \label{Dyn_sys_gamma2}
\end{gather}
According to the slow-roll condition, set $3\bar{H}\gg 1$. Obviously, $(0,0)$ is a singularity. According to Theorem \ref{theory1} in Appendix \ref{definitions_and_properties}, the stability at $(0,0)$ is determined by the matrix
\begin{eqnarray}
      \textbf{A}=\begin{pmatrix}
        0 & 1\\
        -1 & -3\bar{H} \end{pmatrix}, \label{A2}
\end{eqnarray}
whose eigenvalues are $\lambda_1=k_1\ll -1$ and $\lambda_2=k_2\gg -1$, with $k_1k_2=1$, which is exactly the same as the analysis in Sec. \ref{Gamma_constant}.

However, the singularities at infinity are quite different from the those of the system in Eq. (\ref{Dyn_sys_gamma0}). Do the transformation as above: $u={y}/{x}$ and $z={1}/{x}$. Thus,
\begin{eqnarray}
      \begin{cases}
        u'=-(\bar{\Gamma}_2u+z^2+3\bar{H}uz^2+u^2z^2),  \\
        z'=-z^3u,
      \end{cases}\label{Dyn_sys_gamma2_infinity}
\end{eqnarray}
where prime denotes the derivative with respect to $\tau$ and $d\tau=dt/z^2$. The system in Eq. (\ref{Dyn_sys_gamma2_infinity}) is not homogeneous and the singular point $(0,0)$ is called a semihyperbolic singularity, so we need to do more treatment to it. Do transformation $\phi:u=r\bar{u}$, $z=r\bar{z}$. We first perform a transformation in the $z$-direction by setting $\bar{u}=1$, which helps us to study the behavior along $z$-direction. Writing $(u,z)\rightarrow(r,r\bar{z})$, we get
\begin{eqnarray}
      \begin{cases}
        r'=-(\bar{\Gamma}_2r+r^2\bar{z}^2+3\bar{H}r^3\bar{z}^2+r^4\bar{z}^2),  \\
        \bar{z}'=+(\bar{\Gamma}_2\bar{z}+r\bar{z}^3+3\bar{H}r^2\bar{z}^3+2r^3\bar{z}^3).
      \end{cases}\label{Dyn_sys_gamma2_infinity_blowup}
\end{eqnarray}
So the singular point $(0,0)$ is a saddle. Then setting $\bar{u}=-1$, similarly to the analysis above, we find $(0,0)$ is also a saddle (two saddles are located at negative direction and positive direction, respectively). Moreover, the analysis in the $u$-direction becomes simple: according to the semihyperbolic singularity theorem \cite{Dynamic1}, the vector flow on $z$-axis satisfies $u'=-\bar{\Gamma}_2u$. Then, we need to put the vector fields on the planar phase into one point. The topological structure at $(0,0)$ is shown on the second panel in Fig. \ref{blowup_canonical}, which is just a saddle. Such method is called \textit{blowup}. After the calculations above, the singularities $A'$ and $A''$ at infinity are saddles which are located at the $x$-axis (because $u=0$). Similarly, by performing the transformation $v=x/y,\ w=1/y$, we get another two singularities at infinity which are unstable nodes located at $y$-axis.

Comparing Fig.\ref{canonical_gamma0} with Fig.\ref{canonical_gamma2}, we conclude that the topological structures are almost the same at the singular point $(0,0)$, and the structures at infinity are also nearly the same except for their locations. That means the trajectories in the inflationary region $\mathcal{J}$ evolve almost parallel to $y$-axis with the $y\sim0$. This result shows us that the warm inflation model with the dissipative coefficient $\Gamma(\phi)=\Gamma_2 \phi^2$ dramatically increases the possibility of the occurrence of inflation.

\section{\label{Noncanonical_Warm_inflation}Noncanonical Warm Inflation}  
From now on, we will discuss the dynamic properties of noncanonical warm inflation. At beginning of this section, let us start from a condition that is quite analogous with canonical warm inflation as discussed above.
\subsection{\label{K=1}$K(\phi)=1+k\phi^2,\ \alpha>0$}                              
Set the following: $K(\phi)=1+k\phi^2$, $\alpha>0$, $\Gamma(\phi)=\Gamma_2\phi^2$ and $V(\phi)=\frac{1}{2}m^2\phi^2$. Redefine the variables $t\rightarrow t/m$, $\phi\rightarrow M_px$, $\dot{\phi}\rightarrow mM_py$, $H\rightarrow m\bar{H}$, $\Gamma_2\rightarrow (m/M_p^2)\bar{\Gamma}_2$, $k\rightarrow\bar{k}/M_p^2$, and $\alpha\rightarrow\bar{\alpha}/(m^2M_p^2)$. According to Theorem \ref{theorem2} in Appendix \ref{definitions_and_properties}, the expression of such a dynamic dissipative system is written as
\begin{eqnarray} \label{Dyn_nc_K1}
      \begin{cases}
        x'=y(1+\bar{k}x^2+3\bar{\alpha}y^2),  \\
        y'=-x-3\bar{H}y-3\bar{H}\bar{k}x^2y-\bar{\Gamma}_2x^2y-\bar{k}xy^2-3\bar{H}\bar{\alpha}y^3,
      \end{cases}
\end{eqnarray}
where prime denotes the derivative with respect to $\tau$ and $d\tau=dt/(1+\bar{k}x^2+3\bar{\alpha}y^2)$. Using Theorem \ref{theory1} in Appendix \ref{definitions_and_properties} once again, it is easy to get the conclusion that the topological structure of the system in Eq. (\ref{Dyn_nc_K1}) at the original point is the same as the one in the system of Eq. (\ref{Dyn_sys_gamma0}) or the system of Eq. (\ref{Dyn_sys_gamma2}), which are stable nodes, and the topological structures at infinity appear identical to the ones in the system of Eq. (\ref{Dyn_sys_gamma2}), which are semihyperbolic singularities located at the $x$-axis and the $y$-axis. So the global phase portrait is almost the same as the portrait in Fig. \ref{canonical_gamma2}.
\begin{figure*}
  \centering
  \includegraphics[width=6.0in,height=2.7in]{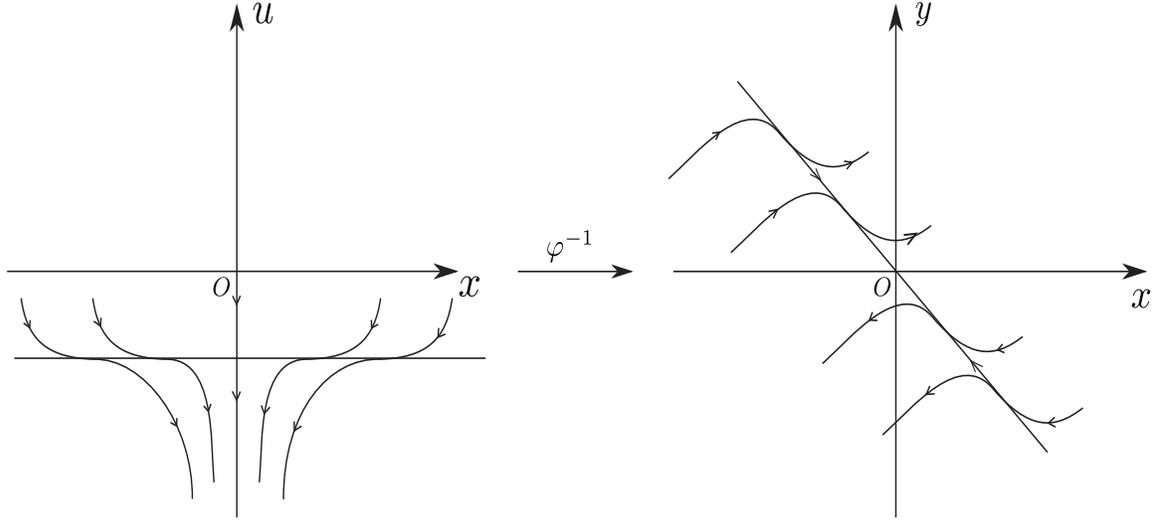}
  \caption{Blowup of the singularity and local phase portrait of Eq. (\ref{dyn_case2}).}\label{blowup_noncanonical_2}
\end{figure*}

\subsection{\label{K_quadric}$K(\phi)=k\phi^2,\ \alpha>0$}                              

Now, let us consider a more complex model. Set the following: $K(\phi)=k\phi^2$, $\alpha>0$, $\Gamma(\phi)=\Gamma_2\phi^2$, and $V(\phi)=\frac{1}{4}\lambda\phi^4$. Redefine the variables $t\rightarrow t/M_p$, $\phi\rightarrow M_px$, $\dot{\phi}\rightarrow M_p^2y$, $H\rightarrow M_p\bar{H}$, $\Gamma_2\rightarrow\bar{\Gamma}_2/M_p$, $k\rightarrow\bar{k}/M_p^2$, $\alpha\rightarrow\bar{\alpha}/M_p^4$, and $V_0=\lambda$; we obtain
\begin{eqnarray} \label{Dyn_nc_K2_V4}
      \begin{cases}
        x'=\bar{k}x^2y+3\bar{\alpha}y^3,  \\
        y'=-3\bar{H}\bar{k}x^2y-\bar{\Gamma}_2x^2y-3\bar{H}\bar{\alpha}y^3-\bar{k}xy^2-x^3.
      \end{cases}
\end{eqnarray}
Next it will be seen that there exist completely different topological structures by choosing different combinations of parameters in the system of Eq. (\ref{Dyn_nc_K2_V4}).

\subsubsection{\label{case1}Case 1}                                                      
Set the following: $\bar{k}=\bar{H}=\bar{\Gamma}=V_0=1$, and $\bar{\alpha}=1/3$. We have
\begin{eqnarray} \label{dyn_case1}
      \begin{cases}
        x'=x^2y+y^3,  \\
        y'=-4x^2y-y^3-xy^2-x^3.
      \end{cases}
\end{eqnarray}
The singularity at the original point $(0,0)$ is a nonelementary one, and the expression of the blowup map is given by
\begin{eqnarray} \label{Briot-Bouquet transformation}
      \varphi:(x,y)\rightarrow(x,ux),
\end{eqnarray}
which is also called Briot-Bouquet transformation that maps $(0,0)$ to the $xOu$ planar. Thus,
\begin{eqnarray} \label{blowup_case1}
      \begin{cases}
        dx/d\eta=ux(1+u^2),  \\
        dy/d\eta=-(u^4+u^3+3u^2+4u+1)\equiv f(u),
      \end{cases}
\end{eqnarray}
where $d\eta=x^2d\tau$. There are two singularities in the system of Eq. (\ref{blowup_case1}): $(0,u_1=-1.33)$ and $(0,u_2=-0.29)$. The linear matrix $\textbf{A}$ reads
\begin{eqnarray}
      \begin{pmatrix}
         u_1(1+u_1^2)<0 & 0 \\
         0 & f'(u_1)>0
      \end{pmatrix}
\end{eqnarray}
and
\begin{eqnarray}
      \begin{pmatrix}
         u_2(1+u_2^2)<0 & 0 \\
         0 & f'(u_2)<0
      \end{pmatrix} .
\end{eqnarray}
So $(0,u_1)$ is a saddle, while $(0,u_2)$ is a stable node whose topological structures on $xOu$ are plotted on the left panel in Fig. \ref{blowup_noncanonical_1}. Then we can obtain the vector fields near $(0,0)$ on the coordinate $xOy$ by map $\varphi^{-1}$, which is drawn on the second panel. Next we turn to the analysis of the singularities at infinity of the system in Eq. (\ref{dyn_case1}).

Using coordinate transformation in Eq. (\ref{Briot-Bouquet transformation}), we have
\begin{eqnarray} \label{infinity_case1}
      \begin{cases}
        du/d\eta=-(u^4+u^3+3u^2+4u+1),  \\
        dz/d\eta=-uz(1+u^2),
      \end{cases}
\end{eqnarray}
where $d\eta=d\tau/z^2$. Similar to the analysis above, we immediately see that singularities $A'$ and $A''$ are saddles that are located at $y=u_2x$, while $B'$ and $B''$ are unstable nodes that are located at $y=u_1x$.

\begin{figure}
  \centering
  \includegraphics[width=3.0in,height=3in]{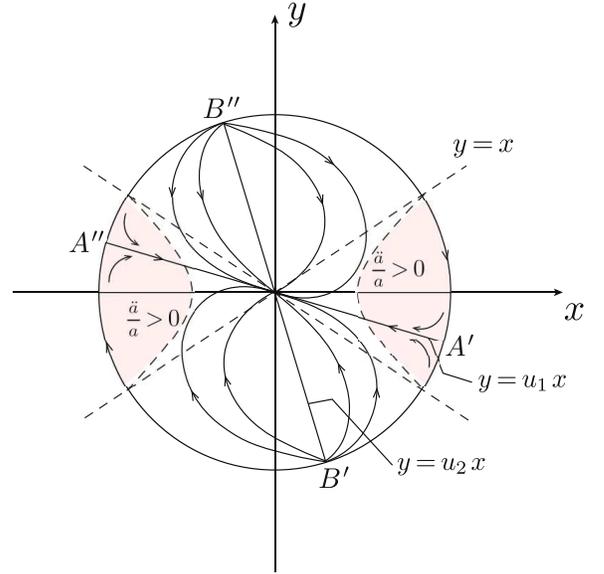}
  \caption{Global phase portrait of the dynamic system in Eq. (\ref{dyn_case1}).}\label{noncanonical_case1}
\end{figure}

Finally, we get the global phase portrait of the dynamic system in Eq. (\ref{dyn_case1}) in Fig. \ref{noncanonical_case1}. From Eq. (\ref{inflationary_region}), we get the region with positive curvature $\mathcal{R}=\{(x,y)|y^2\leqslant x^2\}$, which is the same as the one in Sec. \ref{Gamma_quadric}. In other words, the inflationary regions are almost the same as each other. The direction $y=u_2x$ is repelling, while $y=u_1x$ is an attracting direction. So trajectories evolve leave from the direction $y=u_2x$ and converge along the direction $y=u_1x$. Meanwhile, $\{(x,y)|y=u_1x\}\bigcap\mathcal{J}\neq \emptyset$, which means the orbit evolves for a longer duration in the inflationary region $\mathcal{J}$ compared to the canonical warm inflationary scenario.

\subsubsection{\label{case2}Case 2}                                                      

\begin{figure}
  \centering
  \includegraphics[width=3.0in,height=3in]{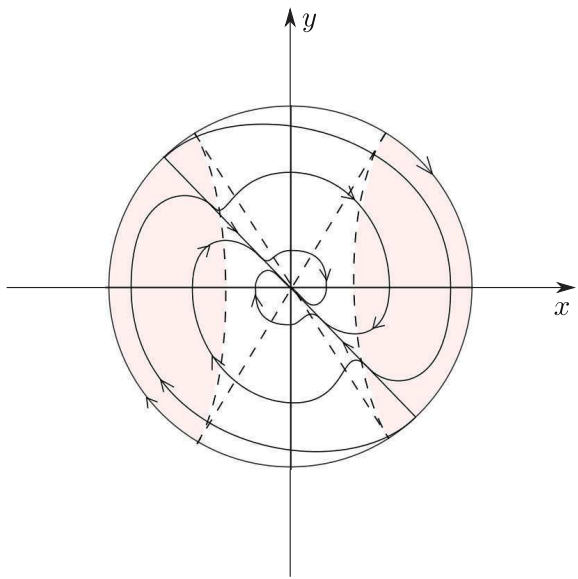}
  \caption{Global phase portrait of the dynamic system in Eq. (\ref{dyn_case2}).}\label{noncanonical_case2}
\end{figure}

Set the following: $\bar{H}=1$, $\bar{\alpha}=1/3$, $\bar{k}=1/4$, $\bar{\Gamma}_2=5/4$, and $V_0=3/2$. The dynamic system is written as
\begin{eqnarray} \label{dyn_case2}
      \begin{cases}
        x'=\frac{1}{4}x^2y+y^3,  \\
        y'=-2x^2y-y^3-\frac{1}{4}xy^2-\frac{3}{2}x^3.
      \end{cases}
\end{eqnarray}
The blowup at the original point is
\begin{eqnarray} \label{blowup_case2}
      \begin{cases}
        dx/d\eta=ux(\frac{1}{4}+u^2),  \\
        du/d\eta=-(u+1)^2(u^2-u+\frac{3}{2}).
      \end{cases}
\end{eqnarray}
The blowup topological structure is a little peculiar: any trajectory crossing $u=u_0=-1$ is tangential to it (see left panel in Fig. \ref{blowup_noncanonical_2}). Transformation $\varphi^{-1}$ means the trajectories near $(0,0)$ cross the line $y=-x$ and are tangential to it (see right panel in Fig. \ref{blowup_noncanonical_2}). However, singularities $A'$ and $A''$ that are located at infinity exhibit a different stability, neither saddles nor nodes, but are called saddle-nodes instead (see Fig. \ref{noncanonical_case2}).

The line $y=u_0x$ lies inside the region $\mathcal{R}=\{(x,y)|y^2\leqslant6x^2\}$ and it must intersect with inflationary region $\mathcal{J}$. As a result, there exist trajectories crossing through the inflationary region, but the condition is weaker than it is in Sec. \ref{case1} because any trajectory will cross the line $y=u_0x$ but it is not a strong attracting direction.

\subsubsection{\label{case3}Case 3}                                                      
\begin{figure}
  \centering
  \includegraphics[width=3.0in,height=3in]{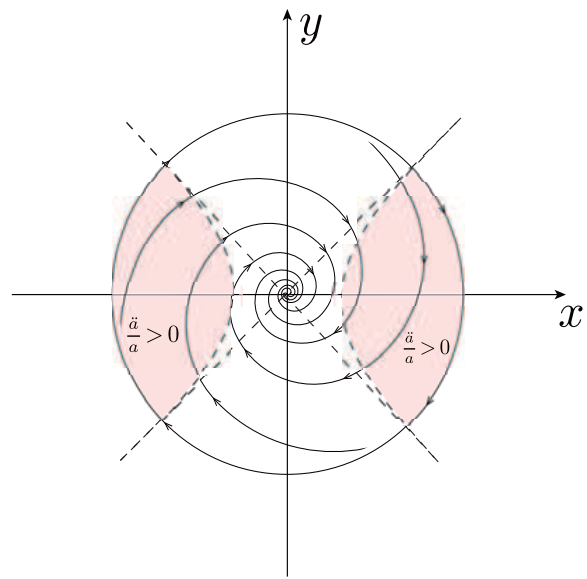}
  \caption{Global phase portrait of the dynamic system in Eq. (\ref{dyn_case3}).}\label{noncanonical_case3}
\end{figure}

Set the following: $\bar{H}=1$, $\bar{\alpha}=1/3$, $\bar{k}=1/4$, $\bar{\Gamma}_2=5/4$, and $V_0=2$. The corresponding system with blowup reads
\begin{eqnarray} \label{dyn_case3}
      \begin{cases}
        x'=\frac{1}{4}x^2y+y^3,  \\
        y'=-2x^2y-y^3-\frac{1}{4}xy^2-2x^3,
      \end{cases}
\end{eqnarray}
with blowup
\begin{eqnarray} \label{blowup_case3}
      \begin{cases}
        dx/d\eta=ux(\frac{1}{4}+u^2),  \\
        du/d\eta=-(u^4+u^3+\frac{1}{2}u^2+2u+2).
      \end{cases}
\end{eqnarray}
Obviously, there is no singular point in Eq. (\ref{blowup_case3}), which means vector fluids will enter the original point on coordinate $xOy$ along no special direction, with the topological structure as something like a stable focus that is plotted on the third panel of Fig. \ref{tological_singularity}. By the transformation in Eq. (\ref{Poincare}), we also see that there is no singularity at infinity as well. The global phase portrait is plotted in Fig. \ref{noncanonical_case3}.

Let us have a brief conclusion of this section. It has been seen that the choice of the noncanonical Lagrangian action determines the behaviors of the dynamic system. In Sec. \ref{K=1}, the topological structure in phase planar is equivalent to the one in canonical warm inflation with the dissipative coefficient $\Gamma=\Gamma_2\phi^2$. In Sec. \ref{K_quadric}, on the other hand, the topological structure is determined by the choice of the combination of parameters. Although we only plot the portraits of three conditions, there also several other global phase portraits like three or four damped directions towards the original point. However, there always exist the phase trajectories that cross the inflationary region, as long as we choose an appropriate initial condition. Finally, based on the accurate analysis above, we reckon that the condition in Sec. \ref{case1} is closer to the physical reality of the inflation model.

\section{\label{other}Other phase trajectory structures of the warm inflation model}    

In the previous sections, we introduced both canonical and noncanonical warm inflationary models whose attractors are all located at the original point. In this section, we will introduce two models with complex topological structures together with different attractors.

\subsection{\label{limit_cycle}Limit cycle}                                        
Regularly, the dissipative coefficient in the warm inflation model is a positive constant or a function larger than zero, and theoretical calculations have also excluded such conditions that dissipative coefficients range less than zero \cite{PhysRevD.58.123508,1475-7516-2009-03-023}. On the other hand, the thermodynamic principle requires a positive dissipative coefficient which is supported by relevant references as well, where a negative one means a violation of the second law of thermodynamics. The second law of thermodynamics in cosmology implies that there are constraints on the effective equation of state of the Universe, in the form of energy conditions, which is obeyed by many known cosmological solutions \cite{0264-9381-15-4-014,1475-7516-2003-05-009}. In semi-de Sitter space, the entropy of a perfect inviscid fluid satisfying the dominant energy condition is proved to be nondecreasing, while, if the fluid is viscous, the generation of entropy ensures that the second law of thermodynamics remains followed \cite{0264-9381-4-6-006,PhysRevLett.84.2072}.

\begin{figure}
  \centering
  \includegraphics[width=3.0in,height=3in]{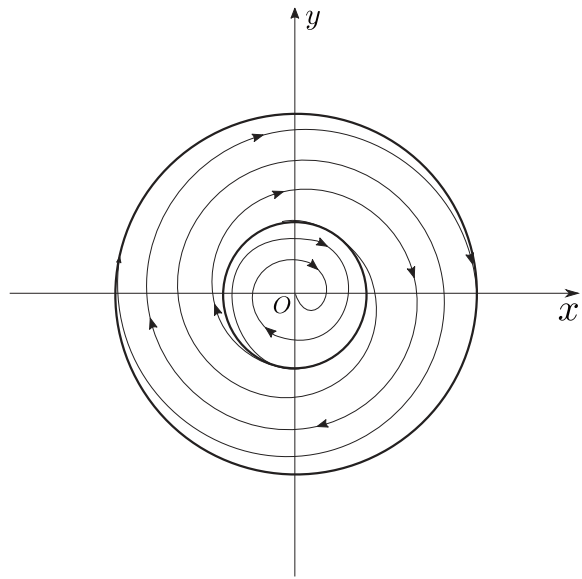}
  \caption{Global phase portrait of the dynamic system in Eq. (\ref{warm_inflation_negativa}).}\label{limit_cycle_portrait}
\end{figure}

Now, we will show this conclusion by dynamical system analysis as well. Redefine variables $t\rightarrow-t$, $\Gamma_2\rightarrow-\bar{\Gamma}_2$ with $\bar{\Gamma}_2>0$, which is similar to the transformation in (\ref{Dyn_sys_gamma2}). The system is a nonlinear oscillation equation
\begin{eqnarray}
      \frac{d^2x}{dt^2}-e(1-rx^2)\frac{dx}{dt}+x=0 \label{oscillation_equation}
\end{eqnarray}
with $e\equiv3\bar{H}$ and $r\equiv\bar{\Gamma}_2/3\bar{H}$. Equation (\ref{oscillation_equation}) is just the famous von der Pol equation. According Theorem \ref{limit_cycle_theorem} in Appendix \ref{definitions_and_properties}, we find the system in Eq. (\ref{oscillation_equation}) satisfies all the conditions, which means the system has a stable limit cycle, or the $\omega$ limit set of the system is a limit cycle. However, we should notice that the result above is obtained under the transformation $t\rightarrow-t$ with the opposite time evolutionary orientation, which means the limit cycle of the initial dynamic equation of warm inflation
\begin{eqnarray}
      \ddot{\phi}+3H\dot{\phi}+\Gamma(\phi)\dot{\phi}+V_\phi=0 \label{warm_inflation_negativa}
\end{eqnarray}
is not stable at all; in other words, the original point $(0,0)$ and infinite region are the attractors of the system in Eq. (\ref{warm_inflation_negativa}) instead of a limit cycle (see Fig. \ref{limit_cycle_portrait}). Obviously, there can not exist the inflationary region in such a model and it represents barely any physical meaning in both theoretical practice and observational practice.

\subsection{\label{bifurcation}Bifurcation}                                     
\begin{figure}
  \centering
  \includegraphics[width=3.0in,height=3in]{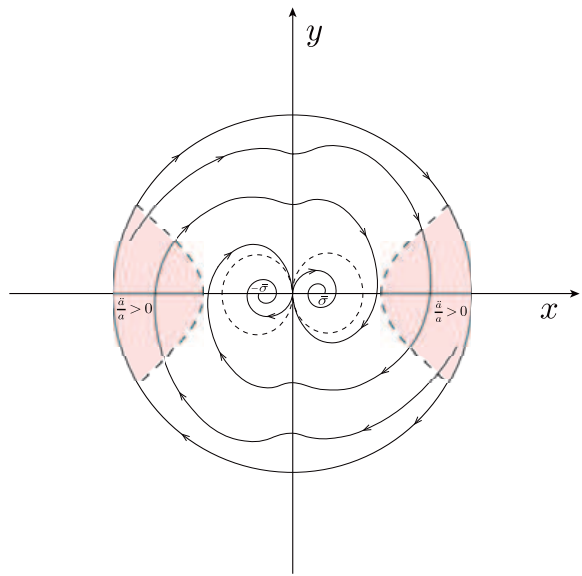}
  \caption{Global phase portrait of the dynamic system in Eq. (\ref{dyn_bifurcation}).}\label{Hopf_bifurcation}
\end{figure}

Consider the potential function $V(\phi)=\frac{1}{4}(\phi^2-\sigma^2)^2$, which is widely used in the models of symmetry breaking in gauge field theory \cite{symmerty_breaking} and reheating theory in cosmology \cite{PhysRevLett.73.3195,PhysRevD.56.3258}. Set the following $K(\phi)=0$, $\bar{\alpha}=1/3$, $\lambda=2$, $\bar{H}=1$, and $\bar{\Gamma}_2=1$, which is similar to the parameters in Sec. \ref{K_quadric}. The dynamic system reads
\begin{eqnarray} \label{dyn_bifurcation}
      \begin{cases}
        x'=y^3,  \\
        y'=-y^3-x^2y-2x(x^2-\bar{\sigma}^2),
      \end{cases}
\end{eqnarray}
where $\bar{\sigma}\equiv\sigma/M_p$. Now, assume $\bar{\sigma}^2<0$ (mathematical respect), the singular point $(0,0)$ is a strong focus. If $\bar{\sigma}=0$, the singular point $(0,0)$ is a week focus that is sensitive to a small perturbation, while, when $\bar{\sigma}^2>0$, there are three singular points on the finite region, $(\pm\bar{\sigma},0)$ and $(0,0)$. We immediately see that $(0,0)$ is an unstable singularity while two stable focuses $(\pm\bar{\sigma},0)$ (see the analysis in Sec. \ref{case3}) arise near the singularity $(0,0)$. Such a singularity is called the Hopf bifurcation point \cite{bifurcation1,bifurcation2}, and its global dynamical phase portrait is plotted in Fig. \ref{Hopf_bifurcation}.

Now, let us have a further discussion. In the reheating model, it suggests that the field oscillates at the bottom of potential function (at $\pm\bar{\sigma}$). However, as discussed previously, some models (dependent on the choice of the parameters) will not oscillate at all, but damp to the bottom directly, which means very few new particles will generate during this period. So reheating appears only under the condition with either a small enough dissipative coefficient (though $\lambda$ is small) or with a large enough $\lambda$ (though the dissipative coefficient is small) \cite{PhysRevD.51.5438,RevModPhys.78.537,PhysRevD.56.3258}.

\section{\label{conclusion}Conclusion and further discussion}                        
In this work we derive a dynamic dissipative system in the phase space of the warm inflationary model and analyze it in both canonical and noncanonical conditions. We first study dynamic systems describing a canonical inflationary dynamic with a different dissipative coefficient. We have also distinguished them by global dynamical analysis in planar phase space: (a) If the dissipative coefficient is a constant, it is a dynamical system just like the standard inflation established initially. The trajectories are attracted along a special direction and exponentially damp to the original point. The singularities at infinity are located at two special directions whose gradients are just the eigenvalues of linear matrix $\textbf{A}$. (b) If the dissipative coefficient is a field dependent function of the quadratic exponential, the global dynamical behavior is quite different. The topological structure at the original point is the same as the one with a constant dissipative coefficient, but the singularities at infinity are located at the $x$-axis and the $y$-axis, which means the trajectories in the inflationary region tend more to the $y$-axis, i.e., $\dot{\phi}\sim0$, which increases the possibility of the occurrence of inflation.

The noncanonical condition is another point we mainly discuss in this work. If we set $K(\phi)=1+k\phi^2,\ \alpha>0$ in noncanonical Lagrangian action, the system exhibits the same dynamic properties as the model of canonical warm inflation. While, if we set $K(\phi)=k\phi^2,\ \alpha>0$, the systems show dramatically different global phase portraits due to the different combinations of parameters (normalized $\bar{H}$ to the unit). From the physical side, we can also reach some interesting conclusions which may have an important meaning in inflationary dynamics. The noncanonical warm inflationary scenarios still have stable attractors of the inflationary phase. For the condition $K(\phi)=k\phi^2,\ \alpha>0$, the inflationary region is almost the same as the region of the canonical, but it keeps a long period during the inflationary phase. As an alternative to the standard inflationary model, the warm inflationary scenario leads the Universe to a moderate temperature so that reheating could be avoided. Our results allow us to reach some conclusions that concentrate more on the debate about reheating.

Besides the usefulness of studying the dynamical behaviors of inflationary system, dynamic analysis can also exclude several inflationary models. Canonical warm inflation with the dissipative coefficient $\Gamma(\phi)=-\Gamma_2\phi^2$ is the system with an unstable limit cycle that all trajectories depart from. There are two attracting points of this dynamic system, the original point and the infinite region, which mean the system is quite sensitive to the initial condition. If the initial point is located outside the limit cycle, the system will evolve to infinity, which cannot occur in the early Universe. The model without self-interaction potential is quite difficult to realize. In such a model, all trajectories distribute nearly parallel to the $y$-axis in phase space; in other words, there exists a quite short period during which a trajectory crosses through the inflationary region. There is another dynamic system that has the Hopf bifurcation structure. This model has potential as the form $V(\phi)=\frac{1}{4}(\phi^2-\sigma^2)^2$ which is widely used in the models of symmetry breaking in gauge field theory and reheating theory in cosmology and dynamic analysis on such models will shed some light on studying the reheating stage just after the inflationary phase.

\acknowledgments
This work was supported by the National Natural Science Foundation of China (Grants No. 11575270, No. 11175019, and No. 11235003).

\appendix

\section{Definitions and properties in the dynamic system}\label{definitions_and_properties}

We first introduce several definitions and properties in the planer dynamic system \cite{Dynamic1,Dynamic2,Dynamic3}. Consider the dynamic system
\begin{eqnarray} \label{D1}
      \begin{cases}
        \dot{x}=P(x,y), \\
        \dot{y}=Q(x,y).
      \end{cases}
\end{eqnarray}
with the boundary condition $P(0,0)=Q(0,0)=0$. Then the system in Eq. (\ref{D1}) becomes
\begin{eqnarray} \label{DD1}
      \begin{cases}
        \dot{x}=\frac{\partial P(0,0)}{\partial x}x+\frac{\partial P(0,0)}{\partial y}y+f(x,y), \\
        \dot{y}=\frac{\partial Q(0,0)}{\partial x}x+\frac{\partial Q(0,0)}{\partial y}y+g(x,y),
      \end{cases}
\end{eqnarray}
where $f(x,y)$ and $g(x,y)$ are the rest part higher than second order. The system in Eq. (\ref{DD1}) can also be written in the form of vectors:
\begin{eqnarray}
      \dot{\textbf{x}}=\textbf{A}\textbf{x}+\textbf{f}(\textbf{x}), \label{DDD}
\end{eqnarray}
where $\textbf{x}=(x,y)'$, $\textbf{f}(\textbf{x})=(f(x,y),g(x,y))'$, and
\begin{eqnarray}
      \textbf{A}=\begin{pmatrix}
        \frac{\partial P}{\partial x}\big|_{(0,0)} & \frac{\partial P}{\partial y}\big|_{(0,0)}\\
        \frac{\partial Q}{\partial x}\big|_{(0,0)} & \frac{\partial Q}{\partial y}\big|_{(0,0)} \end{pmatrix}. \label{A}
\end{eqnarray}
The linear part
\begin{eqnarray}
      \begin{cases}
        \dot{x}=\frac{\partial P(0,0)}{\partial x}x+\frac{\partial P(0,0)}{\partial y}y,\label{Dlinear1}\\
        \dot{y}=\frac{\partial Q(0,0)}{\partial x}x+\frac{\partial Q(0,0)}{\partial y}y, \label{Dlinear2}
      \end{cases}
\end{eqnarray}
i.e.,
\begin{eqnarray}
      \dot{\textbf{x}}=\textbf{A}\textbf{x}, \label{DDDD}
\end{eqnarray}
of the system in Eq. (\ref{DDD}) determines the behavior at elementary singular points.
\newtheorem{definition}{Definition}
\begin{definition}\label{definition_of_singularity} 
If $P(x_0,y_0)=Q(x_0,y_0)=0$, the point $(x_0,y_0)$ is called a \textbf{singularity}. If $\textrm{det}\textbf{A}\neq 0$, the point $(x_0,y_0)$ is called an \textbf{elementary singularity}; if $\textrm{det}\textbf{A}= 0$, the point $(x_0,y_0)$ is called a \textbf{nonelementary singularity}. If $\lambda_1=0$ and $\lambda_2\neq 0$, point $(x_0,y_0)$ is called a \textbf{semihyperbolic singularity}. 
\end{definition}

Let $(0,0)$ be a singular point of the dynamic system, and $\lambda_1$ and $\lambda_2$ be the eigenvalues of the linear part matrix $\textbf{A}$.

\begin{figure}
  \centering
  \includegraphics[width=3.0in,height=3.0in]{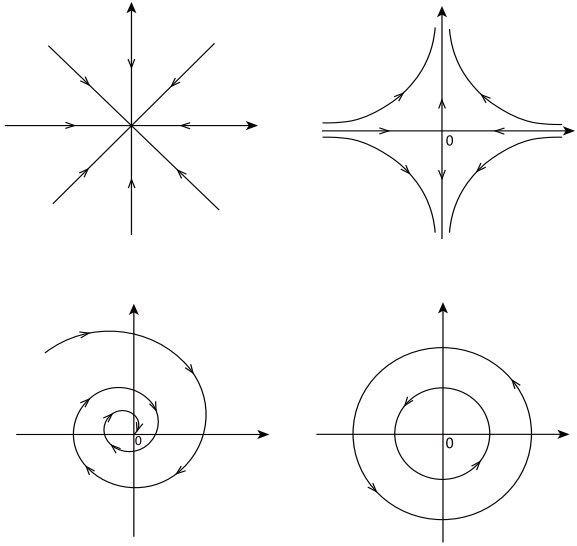}
  \caption{Topological structures of elementary singularities. They are the stable node, saddle, stable focus ($\beta<0$),
  and center ($\beta>0$), respectively.}\label{tological_singularity}
\end{figure}

\begin{definition}\label{definnition_of_elementary_singularity} 
If $\lambda_1<\lambda_2<0(>0)$, $(0,0)$ is a \textbf{stable (unstable) node}. If $\lambda_1\cdot\lambda_2<0$, $(0,0)$ is a \textbf{saddle}. If $\lambda_1=\alpha+\textrm{i}\beta$ and $\lambda_2=\alpha-\textrm{i}\beta$ with $\alpha<0$, $(0,0)$ is a \textbf{stable focus}. If $\lambda_1=\textrm{i}\beta$ and $\lambda_2=-\textrm{i}\beta$, $(0,0)$ is called a \textbf{center} (see Fig. \ref{tological_singularity}). 
\end{definition}

Consider the dynamic systems in Eqs. (\ref{DDD}) and (\ref{DDDD}), we have:

\newtheorem{theorem}{Theorem}
\begin{theorem} \label{theory1}
Assume that $\textbf{f}(\textbf{x})$ is continuous on $\mathbb{R}^2$ and satisfies the Lipschitz condition about $x$, if consistently
\begin{eqnarray}
      \lim_{\textbf{x}\rightarrow\textbf{0}}\frac{\|\textbf{f}(\textbf{x})\|}{\|\textbf{x}\|}=0 \nonumber
\end{eqnarray}
and any eigenvalue of matrix $A$ is not vanished, the stability of the system in Eq. (\ref{DDD}) at $(0,0)$ is the same as the system in Eq. (\ref{DDDD}).
\end{theorem}

Let $X_1$ and $X_2$ be two vector fields on open subsets $D_1$ and $D_2$ on $\mathbb{R}^2$, respectively.

\begin{definition}\label{definition_of_topological_equivalent}
If there exists a homeomorphism $h:D_1\rightarrow D_2$ which maps orbits of $X_1$ to $X_2$ by preventing the orientation, in is said that $X_1$ is a \textbf{topological equivalent to} $X_2$.
\end{definition}

In this paper, the dynamic system of the warm inflation model exhibits a different type which does not follow the same form as Eq. (\ref{D1}); instead, it satisfies the dynamic system
\begin{gather} \label{Dnc1}
      \begin{cases}
        \frac{dx}{dt}=P(x,y), \\
        M(x,y)\frac{dy}{dt}=Q(x,y).
      \end{cases}
\end{gather}
However, is hard to study its stability properties and asymptotic behaviors. We hope that system is topologically equivalent to the dynamic system below:
\begin{eqnarray} \label{Dnc2}
      \begin{cases}
        \frac{dx}{d\tau}=P(x,y)M(x,y), \\
        \frac{dy}{d\tau}=Q(x,y),
      \end{cases}
\end{eqnarray}
where $d\tau=dt/M(x,y)$. The theorem below tells us when they are topologically equivalent to each other.

\begin{theorem} \label{theorem2}
    If $M(x,y)$ is continuous on $\mathbb{R}^2$, $M(x,y)>0$ on $\mathbb{R}^2/\{0\}$ and
    \begin{eqnarray}
      \lim_{t\rightarrow +\infty}M\big(x(t),y(t)\big)=M_0\geqslant 0, \nonumber
\end{eqnarray}
then the system in Eq. (\ref{Dnc1}) is topologically equivalent to the system in Eq. (\ref{Dnc2}).
\end{theorem}

Consider the nonlinear oscillation equation
\begin{eqnarray}
     \frac{d^2x}{dt^2}+f(x)\frac{dx}{dt}+g(x)=0, \label{nonlinear_oscilation_equation}
\end{eqnarray}
where $-g(x)$ is the restoring force and $f(x)$ is the damping force with $f,g\in C(\mathbb{R})$. Integrate Eq. (\ref{nonlinear_oscilation_equation}) on the duration from $0$ to $t$, we have
\begin{eqnarray}
     \frac{dx}{dt}+\int_0^x{f(u)du}+\int_0^t{g(x)dx}=0. \nonumber
\end{eqnarray}
Set the following $y=-\int_0^t{g(x)dx}$ and $F(x)=\int_0^x{f(u)du}$; then we get the Li\'{e}nard equations
\begin{eqnarray} \label{Lienard_equations}
      \begin{cases}
        \frac{dx}{dt}=y-F(x), \\
        \frac{dy}{dt}=-g(x).
      \end{cases}
\end{eqnarray}

\begin{theorem} \label{limit_cycle_theorem}
    Consider Li\'{e}nard equations in Eq. (\ref{Lienard_equations}), if
    \begin{enumerate}
      \item when $x\neq0$, $xg(x)>0$, and
            \begin{eqnarray}
                G(x)=\int_0^x{g(u)du},\ G(\pm \infty)=+\infty; \nonumber
            \end{eqnarray}
      \item when $0<|x|\ll 1$, $xF(x)<0$;
      \item there exist constants $M$ and $k>k'$, such that $F(x)>k$ when $x\geqslant M$ and $F(x)<k'$ when $x\leqslant-M$,
    \end{enumerate}
    then, the system in Eq. (\ref{Lienard_equations}) exists as a stable limit cycle.
\end{theorem}


\end{document}